\documentclass[12pt]{article}
\usepackage{amssymb}
\usepackage{amsmath}
\def\a{\alpha}
\def\b{\beta}

\def\d{\delta}
\def\e{\epsilon}

\def\g{\gamma}
\def\p{\psi}

\def\k{\kappa}

\def\be{\begin{equation}}
\def\ee{\end{equation}}
\def\arr{\begin{array}{rll}}
\def\ea{\end{array}}
\def\bea{\begin{eqnarray}}
\def\eea{\end{eqnarray}}
\def\Z{{\mathbb Z}}
\def\R{{\mathbb R}}

\def\ic{{\rm i}}
\def\eu{{\rm e}}
\def\N2{$N{=}2$}
\def\pa{\partial}

\def\sfrac#1#2{{\textstyle\frac{#1}{#2}}}
\def\>{\rangle}
\def\<{\langle}
\def\+{\dagger}
\def\={\ =\ }
\def\und{\qquad\textrm{and}\qquad}
\begin{document}
\renewcommand{\thefootnote}{\fnsymbol{footnote}}
\begin{titlepage}
\setcounter{page}{0}
\begin{flushright}
LMP-TPU--5/09  \\
ITP-UH--10/09
\end{flushright}
\vskip 1cm
\begin{center}
{\LARGE\bf Harmonic $\mathcal{N}$=2 mechanics}\\
\bigskip
\vskip 1cm
$
\textrm{\Large Anton Galajinsky\ }^{a} \quad \textrm{\Large and} \quad
\textrm{\Large Olaf Lechtenfeld\ }^{b}
$
\vskip 0.7cm
${}^{a}$ {\it
Laboratory of Mathematical Physics, Tomsk Polytechnic University, \\
634050 Tomsk, Lenin Ave. 30, Russian Federation} \\
{Email: galajin@mph.phtd.tpu.ru}
\vskip 0.4cm
${}^{b}$ {\it
Institut f\"ur Theoretische Physik, Leibniz Universit\"at Hannover,\\
Appelstrasse 2, D-30167 Hannover, Germany} \\
{Email: lechtenf@itp.uni-hannover.de}
\vskip 0.7cm

\end{center}
\vskip 1cm
\begin{abstract} \noindent
$\mathcal{N}{=}2$ superconformal many-body quantum mechanics in arbitrary 
dimensions is governed by a single scalar prepotential which determines
the bosonic potential and the boson-fermion couplings. We present a special
class of such models, for which the bosonic potential is absent.
They are classified by homogeneous harmonic functions subject to physical
symmetry requirements, such as translation, rotation and permutation 
invariance. The central charge is naturally quantized. We provide
some examples for systems of identical particles in any dimension.
\end{abstract}

\vskip 1cm
\noindent
PACS numbers: 03.65.-w, 11.30.Pb, 11.30.-j

\vskip 0.5cm

\noindent
Keywords: supersymmetric quantum mechanics

\end{titlepage}

\renewcommand{\thefootnote}{\arabic{footnote}}
\setcounter{footnote}0

\section{Introduction}

$\mathcal{N}{=}4$ superconformal many-body quantum mechanics in one dimension
is governed by two scalar prepotentials $U$ and $F$ which obey a coupled set of
partial differential equations.
While $U$ may vanish, $F$ always takes nonzero values.
Recent studies in \cite{wyl}--\cite{glp2} 
(for related developments see \cite{bgk}--\cite{fil2})
revealed an interesting link between $\mathcal{N}{=}4$ quantum mechanics and 
the WDVV equation~\cite{w,dvv} which plays an important role in $d{=}2$ 
topological field theory~\cite{w,dvv} and $\mathcal{N}{=}2$ supersymmetric 
Yang-Mills theory~\cite{mmm}. Because the WDVV equation underlies a potential 
deformation of a Fr\"obenius algebra~\cite{dub}, it relates
$\mathcal{N}{=}4$ mechanics with Fr\"obenius manifolds.
All $\mathcal{N}{=}4$ models with a nontrivial $U$ constructed so far are
based on the root systems of simple Lie algebras or Coxeter reflection groups.

A peculiar feature of $\mathcal{N}{=}4$ mechanics concerns the center-of-mass
coordinate. Although it decouples from the relative particle motion, its
nonzero $F$~prepotential generates an inverse-square potential for the
center-of-mass motion, thus breaking translation invariance.
If this is unwanted, one must give up $\mathcal{N}{=}4$ and soften the model
to an $\mathcal{N}{=}2$ system, which is ruled by the prepotential~$U$ alone
\cite{fm}. Our interest in $\mathcal{N}{=}2$ mechanics is also motivated
by the desire to go beyond $d{=}1$ and to construct new exactly solvable 
many-body models in higher dimensions and to explore novel correlations 
(see e.g.~\cite{gm} and references therein).
It is natural to expect that $d{>}1$, $\mathcal{N}{=}2$ superconformal 
many-body models will provide new insight into the nonrelativistic version 
of the AdS/CFT correspondence which has currently sparked substantial interest.

A minimal extension of the Galilei algebra by the dilatation and special 
conformal generators is known in the literature as the Schr\"odinger algebra. 
A conformal extension obtained by contracting the relativistic conformal 
$so(d{+}1,2)$ algebra gives an even larger algebra which goes under the name of
conformal Galilei algebra (for a recent discussion and further references see 
e.g.~\cite{dh}). Because the conformal Galilei algebra requires vanishing mass,
the Schr\"odinger algebra has a better prospect for quantum mechanical 
applications. Since the translations are part of the Schr\"odinger algebra, 
$\mathcal{N}{=}2$ interacting many-body quantum mechanics is likely 
to be the maximal superextension feasible in higher dimensions.

The purpose of this paper is to reconsider the construction of 
$\mathcal{N}{=}2$ $n$-particle quantum mechanics in $d$~dimensions and to 
exhibit a new special class of models determined by a single harmonic function.
These $(n,d)$~models are characterized by the absence of bosonic interactions, 
yet retain (quantum) boson-fermion couplings. 
They are classified by the homogeneous harmonic functions on~$\R^{nd}$ subject 
to physical symmetry requirements (Euclidean and permutation invariance)
and quantize the central charge of the $\mathcal{N}{=}2$ algebra.

In Section 2 we recall the conventional framework for formulating 
$\mathcal{N}{=}2$ many-body models in one dimension and explore the
hitherto unexploited possibility of purely boson-fermion couplings.
We show how the Laplace equation arises, explain the central charge 
quantization and discover solutions related to Lie-algebra root systems.

In Section 3 the analysis is extended beyond one dimension. 
It is shown that the role of the Laplace equation persists in higher 
dimensions, but the prepotential is further constrained by Euclidean
invariance in~$\R^d$, as part of the $\mathcal{N}{=}2$ Schr\"odinger 
supersymmetry. We finally present a one-parameter family of $(n,d)$
models as well as a particular $(n,n{-}1)$~system, both being invariant
under particle permutations. Conclusions follow.

\section{Special $\mathcal{N}$=2 mechanics}

The conventional representation of the $d{=}1$, $\mathcal{N}{=}2$ 
superconformal algebra on the phase space of $n$ identical particles 
(with unit mass) is provided by a single prepotential $U(x_1,\dots,x_n)$ 
which gives rise to the operators~\cite{glp}~\footnote{We work in the
standard coordinate representation, $p_i={-}\ic\frac{\pa}{\pa x_i}$,
$[x_i, p_j]=\ic\d_{ij}$, and put $\hbar{=}1$. 
The fermionic operators are mutually conjugate via
${(\p_i)}^{\dagger}{=}\bar\p_i$ and obey the anticommutation relations 
$\{\p_i,\p_j\}{=}0$, $\{\bar\p_i,\bar\p_j\}{=}0$, $\{\p_i,\bar\p_j\}{=}\d_{ij}$.
The $t$-dependent pieces in the generators are kept explicit so as to have
a direct link to the classical theory. Throughout the paper summation over 
repeated indices is understood.}
\bea\label{g}
&&
H\=\sfrac 12 p_i p_i +\sfrac 12 \partial_i U(x) \partial_i U(x) 
-\partial_i \partial_j U(x) \<\p_i \bar\p_j\>, \quad \qquad
J\=\<\p_i \bar\p_i\>,
\nonumber\\[2pt]
&&
D\=tH-\sfrac{1}{4} (x_i p_i +p_i x_i), \quad \qquad
K\=-t^2 H+2t D+\sfrac{1}{2} x_i x_i,
\nonumber\\[2pt]
&&
Q\=\p_i (p_i+\ic \partial_i U(x)), \quad \quad \qquad ~~
\bar Q\= \bar\p_i (p_i-\ic \partial_i U(x)),
\nonumber\\[2pt]
&&
S\=x_i \p_i-t Q, \qquad \quad \quad \quad \qquad ~~
\bar S\=x_i \bar\p_i-t \bar Q,
\eea
where the symbol $\<\ldots\>$ stands for symmetric (or Weyl) ordering of the
fermions. The operators $H$, $D$ and $K$ generate time translations, 
dilatations and special conformal transformations, respectively, while
$Q$ and $\bar Q$ are supersymmetry generators, and $S$ and $\bar S$ generate 
superconformal transformations. The U(1) R-symmetry transformation
generated by~$J$ affects only the fermions. Note that
the prepotential $U(x)$ is defined up to an additive constant.

The operators (\ref{g}) obey the (anti)commutation relations of 
the $d{=}1$, $\mathcal{N}{=}2$ superconformal algebra with central charge~$C$
(Hermitian conjugates are omitted)
\bea\label{algebra}
&&
[H,D]=\ic H, \quad \quad
[K,D]=-\ic K, \quad ~
[Q,D]=\sfrac{\ic}{2} Q, \quad \quad ~~
[S,D]=-\sfrac{\ic}{2} S,
\nonumber\\[4pt]
&&
[Q,J]=- Q, \quad \quad
[S,J]=-S, \quad \quad ~
[H,K]=2\ic D, \quad \quad ~
[Q,K]=-\ic S,
\nonumber\\[4pt]
&&
[S,H]=\ic Q, \qquad
\{Q,\bar Q\}=2H, \qquad
\{S,\bar S\}=2K, \qquad ~
\{Q,\bar S\}=-2D-\ic J-\ic C,
\eea
provided the prepotential satisfies the linear
partial differential equation
\be\label{str}
x_i \partial_i U(x)\=-C\ .
\ee
The general solution to (\ref{str}) reads
\be\label{gs}
U(x)\=-C \ln |x_1|\ +\ \Lambda(\sfrac{x_i}{x_j})\ ,
\ee
where $\Lambda(\sfrac{x_i}{x_j})$ is a function 
of the coordinate ratios $\sfrac{x_i}{x_j}$ for $i<j$.

In order to extract a class of reasonable models from the infinity of 
$\mathcal{N}{=}2$ systems encoded in the general solution~(\ref{gs}), 
one can impose additional restrictions like permutation symmetry, 
translation invariance etc..
Another option is to start with a specific bosonic theory,
\be\label{bos}
H_B\=\sfrac 12 p_i p_i +V(x) \qquad\textrm{with}\qquad
(x_i\partial_i+2) V(x)\=0\ ,
\ee
and then solve the Hamilton-Jacobi equation
\be\label{confp}
\partial_i U(x)\,\partial_i U(x)\=2\,V(x)
\ee
for the given potential $-V$ and zero energy.  
Each solution~$U$ yields an $\mathcal{N}{=}2$ superconformal extension 
of the original model (\ref{bos}).
In particular, in this way one can treat quantum integrable many-body
models related to simple Lie algebras, the prominent example being 
the $\mathcal{N}{=}2$ Calogero model \cite{fm} (see also~\cite{glp}). 

Among the many possible bosonic starting points, there exist special
bosonic potentials~$V$ which can be absorbed into a reordering of the fermions.
Since a deviation from Weyl ordering produces a term proportional
to $\partial_i\partial_jU$ in~$H$, this property translates to the condition
\be\label{u}
\partial_i U(x)\,\partial_i U(x)\ +\ \k\,\partial_i \partial_i U(x) \=0
\ee
for some real parameter~$\k$ of order~$\hbar$. Note that this forces
$U$ to be of order~$\hbar$ as well, so that these models are classically free. 
The value of~$\kappa$ quantifies the deviation from Weyl ordering and takes 
unit ($\hbar$) value for normal ordering.
If (\ref{u}) can be solved, then the Hamiltonian may be brought to the form
\be
H\=\sfrac 12 p_i p_i\ -\ \partial_i \partial_j U(x) :\!\p_i \bar\p_j\!:_\kappa
\ee
for a suitable fermionic ordering prescription,
so that the interaction contains only boson-fermion couplings.
We now describe a class of solutions to~(\ref{u}) 
with quantized central charge.

The conditions (\ref{str}) and (\ref{u}) simplify under the substitution
\be\label{GG}
U(x)=\k\,\ln G(x) 
\qquad\textrm{to}\qquad
(x_i \partial_i+\sfrac{C}{\k})\,G(x)=0 \und \partial_i \partial_i G(x)=0\ ,
\ee
so that $G(x)$ is a harmonic homogeneous function of degree 
$\ell:=-\sfrac{C}{\k}$ in~$\R^n$.
Such functions are single-valued only for $\ell\in\Z$ and regular at the
origin $x_1=x_2=\ldots=x_n=0$ for $\ell\ge0$. These conditions quantize
the central charge in units of~$\k$,
\be
C \= -\ell\,\k \qquad\textrm{with}\quad \ell=0,1,2,\ldots\ ,
\ee
and restrict the prepotential to~\footnote{
If singular behavior is admitted at coincidence loci $x_i{=}x_j$,
more general solutions appear.}
\be
G_\ell(x)\=(x_1^2{+}\ldots{+}x_n^2)^{\sfrac\ell2}\ Y_\ell(\textrm{angles})\ ,
\ee
where $Y_\ell$ is a linear combination of $S^{n-1}$ spherical harmonics
for spin~$\ell$. Note that linear combinations of $G_\ell$ are forbidden 
by the homogeneity condition (\ref{GG}). 

Each value of $\ell$ and choice of~$Y_\ell$ produces a special 
$\mathcal{N}{=}2$ many-body quantum system.
The demand for permutation invariance or translation invariance puts
restrictions on~$Y_\ell$, which can be solved. 
For illustration, we consider a solution related to
the positive roots $\{\alpha\}$ of a simple Lie algebra,
\be\label{la}
G(x)\=\prod_{\alpha} (\alpha x)\ .
\ee
In this case $\ell$ equals the number of positive roots. 
That (\ref{la}) solves Laplace's equation is verified
with the use of the same root identities which were previously applied 
in~\cite{glp2} for solving the WDVV equation 
(see section 6~in~\cite{glp2} for more details).
Permutation and translation invariance it achieved for the $A_n$ root systems,
$\{(\a x)\}=\{x_i{-}x_j\,|\,1{\le}i{<}j{\le}n{+}1\}$.
The interaction potential for these models reads
\be
V_{int}\=\sum_\a \frac{(\a\p)(\a\bar\p)}{{(\a x)}^2}\ .
\ee

\section{Special $\mathcal{N}$=2 models in arbitrary dimension}

We proceed to the construction of $\mathcal{N}{=}2$ models in dimensions~$d$
greater than one. At the algebraic level, extra dimensions come with
additional generators corresponding to spatial translations $P^\a$, spatial
rotations $M^{\a\b}$, Galilei boosts $K^\a$ and super Galilei transformations
$L^\a$ and $\bar L^\a$ with $\a,\b=1,\dots,d$. It is assumed that $L^\a$ and 
$\bar L^\a$ are Hermitian conjugates of each other.
The set of generators $\{H,D,K,P^\a,K^\a,M^{\a\b}\}$ spans a subalgebra known 
as the Schr\"odinger algebra.
In what follows, Greek letters are reserved for spatial indices while 
Latin indices label identical particles of unit mass.

Apart from the structure relations (\ref{algebra}), which persist in higher 
dimensions, the non-vanishing (anti)commutation relations of the
$\mathcal{N}{=}2$ Schr\"odinger superalgebra include 
(Hermitian conjugates are omitted)
\bea\label{algebra1}
&&
[H,K^\a]=-\ic P^\a,\qquad [D,K^\a]= \sfrac{\ic}{2} K^\a, \qquad
[K,P^\a]= \ic K^\a,\qquad [D,P^\a]=-\sfrac{\ic}{2} P^\a,
\nonumber\\[2pt]
&&
[M^{\a\b},L^\g]=\ic(\d^{\a\g} L^\b-\d^{\b\g} L^\a), \qquad \qquad \quad ~ 
[Q,K^\a]=-\ic L^\a , \quad ~
[K^\a,P^\b]=\ic\d^{\a\b} M,
\nonumber\\[3pt]
&&
[M^{\a\b},P^\g]=\ic(\d^{\a\g} P^\b-\d^{\b\g} P^\a), \qquad \qquad \quad  
\{Q,\bar L^\a \}=P^\a, \qquad 
\{L^\a,\bar L^\b\}=\delta^{\a\b} Z,
\nonumber\\[3pt]
&&
[M^{\a\b},K^\g]=\ic(\d^{\a\g} K^\b-\d^{\b\g} K^\a), \qquad \quad \quad ~ ~ 
[S,P^\a]=\ic L^\a, \qquad ~
\{S,\bar L^\a\}=K^\a,
\nonumber\\[3pt]
&&
[M^{\a\b},M^{\g\d}]=\ic
(\d^{\a\g}M^{\b\d}+\d^{\b\d}M^{\a\g}-\d^{\b\g}M^{\a\d}-\d^{\a\d}M^{\b\g}), 
\qquad ~ ~ ~  [J,L^\a]=L^\a,
\eea
where $M$ and $Z$ are the central charges.

In order to build a quantum mechanical representation of this algebra, 
one introduces bosonic operators $x_i^\a$, $p_i^\a$ and
fermionic operators $\p_i^\a$, $\bar\p^\a_i$,   
which obey the (anti)commutation relations
\be
[x_i^\a,p_j^\b]\=\ic\delta^{\a\b} \d_{ij} \und
\{\p_i^\a,{\bar\p}_j^\b \}\=\delta^{\a\b} \delta_{ij}\ .
\ee
The fermionic operators are related by Hermitian conjugation, 
i.e.~${(\psi_i^\a)}^{\dagger}={\bar\psi}_i^\a$.
A representation of the superalgebra~(\ref{algebra1}) can then be constructed 
in terms of a single prepotential $U(x)$ by analogy with the one-dimensional 
case,
\bea\label{gener}
&&
Q=\p^\a_i (p^\a_i+\ic \partial_{\a i} U(x)), \qquad
\bar Q= \bar\p^\a_i (p^\a_i-\ic \partial_{\a i} U(x)), \qquad   
L^\a={\textstyle\sum_i} \p^\a_i, \qquad 
{\bar L}^\a={\textstyle\sum_i} \bar\p^\a_i,
\nonumber\\[2pt]
&&
S=x^\a_i \p^\a_i-t Q, \qquad \qquad \quad
\bar S=x^\a_i \bar\p^\a_i -t \bar Q, \qquad \qquad \qquad 
J=\<\p^\a_i \bar\p^\a_i\>, \ \ \quad 
P^\a={\textstyle\sum_i} p^\a_i,
\nonumber\\[2pt]
&&
M^{\a\b}=(x^\a_i p^\b_i-x^\b_i p^\a_i) 
-\ic\<\p^\a_i \bar\p^\b_i-\p^\b_i \bar\p^\a_i\>, \qquad \qquad \qquad \quad 
K^\a={\textstyle\sum_i} x^\a_i-t P^\a,
\nonumber\\[2pt]
&&
C=-t^2 H+2t D+\sfrac{1}{2} x^\a_i x^\a_i, \qquad \qquad \quad 
D=tH-\sfrac{1}{4}(x^\a_i p^\a_i+ p^\a_i x^\a_i),
\nonumber\\[2pt]
&&
H=\sfrac{1}{2}p^\a_i p^\a_i+\sfrac12\partial_{\a i}U(x)\partial_{\a i}U(x)-
\partial_{\a i}\partial_{\b j}U(x)\<\p^\a_i \bar\p^\b_j\>,
\eea
where we abbreviated $\partial_{\a i}=\frac{\partial}{\partial x^\a_i}$. This 
representation fixes the values of the two central charges to $Z=M=n$.\footnote{
For particles of mass $m$ one has $Z=M=nm$.} 
The commutation relations of the $\mathcal{N}{=}2$ Schr\"odinger superalgebra 
(\ref{algebra1}) constrain the prepotential to obey a set of partial 
differential equations,
\bea\label{str1}
\bigl(x_i^\a \pa_{\b i}-x_i^\b \pa_{\a i}\bigr)U(x)\=0\ , \qquad
{\textstyle\sum_i}\pa_{\a i} U(x)\=0\ ,\qquad
x_i^\a \pa_{\a i} U(x)\=-C\ .
\eea
The first two restrictions in (\ref{str1}) come from rotation and translation 
invariance, while the last one is responsible for conformal symmetry.

Like in one dimension, we would like to absorb the bosonic potential 
$V=\sfrac12\pa_{\a i}U\pa_{\a i}U$ into a reordering of the fermions.
The condition for this option generalizes our principal equation (\ref{u}) to
\be
\pa_{\a i}U(x)\pa_{\a i}U(x) + \k\,\pa_{\a i}\pa_{\a i}U(x)\= 0\ .
\ee
Introducing $G(x)$ as in (\ref{GG}) one gets 
\be\label{Gd}
(x_i^\a \pa_{\a i}+\sfrac{C}{\k})\,G(x)\=0 \und
\pa_{\a i}\pa_{\a i} G(x)\=0
\ee
besides translation and rotation invariance for~$G(x)$.

Formally, the $n$-particle model in $d$ dimensions is just a special 
$nd$-particle model in one dimension. However, the physical symmetry 
requirement is different: We want the potential to be invariant under
permutations of the $n$ particle labels only, and not under permutations 
of all $nd$ labels. The translation and rotation invariance, on the other 
hand, is more restrictive in $d$~dimensions, but this may be dealt with by 
passing to a set of SO($d$) invariants built from relative coordinates.
We shall see that for $d{>}1$ it is possible to construct physically 
acceptable $(n,d)$ models for identical particles.

Like in one dimension, we consider prepotentials~$G$ which are regular at 
the origin $x_i^\a\,{=}\,0$. We don't know how to write down the most general
rotation and translation invariant harmonic function, but let us present two
classes of examples.
In order to take into account translation and rotation invariance, 
we switch to the relative coordinates $r^\a_{ij}=x^\a_i-x^\a_j$
and form SO($d$) scalars $(r_{ij},r_{kl})$ and 
$\epsilon_{\a_1\dots\a_d}r^{\a_1}_{i_1j_1}\dots r^{\a_d}_{i_dj_d}$ from them.
Here, $(~,~)$ and $\epsilon_{\a_1\dots\a_d}$ denote the Euclidean scalar 
product and the Levi-Civita tensor, respectively, in $\R^d$.
It should be kept in mind that these building blocks are not independent, 
e.g.~the triangle rule $r^\a_{ij}+r^\a_{jk}+r^\a_{ki}=0$ implies that
\be\label{dep}
(r_{ij},r_{jk})+(r_{jk},r_{ki})+(r_{ki},r_{ij})=-\sfrac12\,[
(r_{ij},r_{ij})+(r_{jk},r_{jk})+(r_{ki},r_{ki})] \quad \textrm{(no sum)}\ .
\ee

Our first example is a homogeneous and permutation invariant polynomial 
of fourth order (thus $C=-4\kappa$),
\be\label{g1}
G(x)=\a\sum_{i<j,k}^n (r_{ik},r_{kj})^2\ +\
\b\sum_{i<j,k}^n (r_{ik},r_{ik})(r_{kj},r_{kj})\ +\
\g\sum_{i<j}^n (r_{ij},r_{ij})^2 
\ee
with free parameters $\a$, $\b$ and~$\g$. Computing 
\be
\begin{aligned}
\pa_{\a m} \pa_{\a m}\ (r_{ik},r_{kj})^2 &\=
4\,(r_{ik},r_{ik})+4\,(r_{kj},r_{kj})-4(d{+}1)(r_{ik},r_{kj})\ ,\\
\pa_{\a m} \pa_{\a m}\ (r_{ik},r_{ik})(r_{kj},r_{kj})&\=
4d\,(r_{ik},r_{ik})+4d\,(r_{kj},r_{kj})-8\,(r_{ik},r_{kj})\ ,\\
\pa_{\a m} \pa_{\a m}\ (r_{ij},r_{ij})^2 &\=
8(d{+}2)(r_{ij},r_{ij})
\end{aligned}
\ee
with sums over $m$ only,
and employing the identity 
\be\label{i}
\sum_{i<j,k}^n (r_{ik},r_{kj}) \= 
\sfrac{2-n}{2}\,\sum_{i<j}^n (r_{ij},r_{ij})
\ee
following from (\ref{dep}), one arrives at
\be
\pa_{\a m} \pa_{\a m} G(x) \= \delta\ \sum_{i<j}^n (r_{ij},r_{ij})\ ,
\ee
with $\delta$ being a linear expression in $\a$, $\b$ and~$\g$.
Therefore, solving (\ref{Gd}) amounts to putting $\delta=0$, which is
\be
(n{-}2)(d{+}5)\,\a\ +\ (n{-}2)(4d{+}2)\,\b\ +\ (4d{+}8)\,\g \= 0\ .
\ee
Since the scale of~$G(x)$ is irrelevant, this linear relation leaves
a one-parameter family~(\ref{g1}) of $(n,d)$ prepotentials, for
$d{>}1$ and $n{\ge}2$. The formulae also work for $d{=}1$, but
produce $G(x)\equiv0$.

Viewing the three particle labels $i,j,k$ in (\ref{g1}) as the vertices
of a triangle, this prepotential appears to be constructed in terms of
triangle areas and edge lengths. 
This suggests to construct other prepotentials in terms of generalized volumes.
The simplest such situation, specific to $d=n{-}1$ dimensions, provides
our second example,
\be\label{ep}
G(x)\=\e_{\a_1\dots\a_{n-1}}r^{\a_1}_{12}r^{\a_2}_{13}\dots r^{\a_{n-1}}_{1n}
\ .
\ee
This homogeneous polynomial of degree $n{-}1$ measures the volume of the
simplex spanned by the $n=d{+}1$ particle locations and is naturally
permutation invariant (up to an irrelevant sign). It trivially solves the 
Laplace equation since each vector~$x^\a_i$ occurs at most linearly 
in~(\ref{ep}). Hence, this example describes a valid $(n,n{-}1)$ particle model.

\section{Conclusions}

We have constructed new interacting $\mathcal{N}{=}2$ many-body quantum
mechanics of a special kind: the bosonic potential is absent, but interaction
takes place through boson-fermion couplings alone. These couplings are
governed by a prepotential~$G=\eu^{U/\kappa}$ which only has to be harmonic 
and homogeneous. The central charge (in the $\mathcal{N}{=}2$ superconformal 
algebra) is given by the degree of~$G$ and therefore naturally quantized.
By changing the fermionic ordering prescription, one may generate also
a particular bosonic potential which is purely quantum.

In $d{=}1$, the admissible prepotentials include models built from the 
positive roots of simple Lie algebras. The $A_n$ root systems yield
translation-invariant models of identical particles. 
In dimensions $d{>}1$, we provided a general framework with
$\mathcal{N}{=}2$ Schr\"odinger supersymmetry and gave two example models, 
one for generic $(n,d)$ with a free parameter and another one for $d=n{-}1$.

Finally, let us discuss possible further developments of this work. 
The quantization of the central charge may be weakened by letting the
particle coordinates parametrize a cone rather than $\R^n$. The freedom
of a deficit angle around the singularity allows for more general harmonic 
functions and therefore other $\mathcal{N}{=}2$ models.
In the higher-dimensional situation, 
our examples were not the most general ones.
A physical classification needs an understanding of all homogeneous harmonic 
functions on $\R^{nd}$ invariant under the $n!$ permutations of the particle 
labels and under the rigid translations and rotations of~$\R^d$. It would be 
interesting to learn how the root-system solutions fit into such a scheme.

\vspace{0.5cm}

\noindent{\bf Acknowledgements}\\
\noindent
This work was supported in part by DFG grant 436 RUS 113/669/0-3,
RF Presidential grant MD-2590.2008.2 and RFBR grant 09-02-91349.
O.L.~thanks D.~Fairlie for help with~(\ref{u}).

\end{document}